\documentclass[12pt]{article} 

\usepackage[utf8]{inputenc} 

\usepackage{geometry} 
\usepackage{graphicx} 

\usepackage{booktabs} 
\usepackage{array} 
\usepackage{paralist} 
\usepackage{verbatim} 
\usepackage{subfig} 
\usepackage{color}
\usepackage{crayola}
\usepackage{amsmath}
\usepackage{kbordermatrix}
\usepackage{algpseudocode} 
\usepackage{hyperref}
\usepackage{algorithm}
\usepackage{fancyhdr} 
\usepackage{sectsty}
\allsectionsfont{\sffamily\mdseries\upshape} 

\usepackage[nottoc,notlof,notlot]{tocbibind} 
\usepackage[titles,subfigure]{tocloft} 

\newcommand{\keyw}[1]{\textcolor{red}{\emph{#1}}}

\newcommand{\clock}{\count254=\time \divide\count254 by 60
 \count255=\count254 \multiply\count255 by -60
 \advance\count255 by \time
 \ifnum\count254<10 0\fi\number\count254\,:\,%
 \ifnum\count255<10 0\fi\number\count255}

\newif\ifdraft \newif\ifblind
\draftfalse 
\blindfalse
\ifblind\else\fi

\title{\textbf{On weighted two-mode network projections}}
\ifblind\author{Author\\ Institution\\ e-mail}\else
\author{Vladimir Batagelj\\IMFM Ljubljana and IAM UP Koper\\ 
e-mail: \texttt{vladimir.batagelj@fmf.uni-lj.si}\\ ORCID: 0000-0002-0240-9446}
\fi
\ifdraft\date{\today\  at \clock}\else\date{}\fi
\ifblind\else
\hypersetup{pdfauthor={V. Batagelj}}
\hypersetup{pdftitle={On weighted two-mode network projections}}
\fi

\begin{document}
\maketitle
\begin{abstract}
The standard and fractional projections are extended from binary two-mode networks to weighted two-mode networks. 
Some interesting properties of the extended projections are proved.\smallskip

\end{abstract}
\section{Introduction}

In the paper \cite{soviet} \ifblind\else we studied\fi{} the collaboration (co-authorship) between scientists from different post-Soviet countries\ifblind{} was studied\fi. We decided to repeat the study on the European countries. It turned out that there are different ways how we can define a network describing the co-authorship collaboration between countries. Some options are discussed in this paper.

Most of the bibliometric networks are obtained by a projection of a non-weighted network represented by a binary matrix. For example from the authorship network $W\!\!A$ describing the authorship relation of the set of works  (papers, books, reports, etc.) $W$ by the authors from the sets $A$. It is represented by a matrix $\mathbf{WA} = [wa[w,a]]$ where $wa[w,a] = 1$  iff $a$ is an author of the work $w$ and $0$ otherwise. We get the co-authorship (counting) network $Co_A$ determined by the projection
\[ \mathbf{Co_A} = \mathbf{WA}^T \cdot \mathbf{WA}\]
As we know \cite{onbib}
\begin{itemize}
\item For $a\ne b$, $co_A[a,b] =$ number of works co-authored by authors $a$ and $b$.
\item $co_A[a,a] =$ number of works from $W$ written by the author $a$.
\item The works with a large number of coauthors are "overrepresented" in the network $Co_A$ -- for example, the co-authorship of authors of a paper with 2 authors counts the same as the co-authorship between any pair of authors of the paper with 1000 co-authors; a paper with 1000 co-authors adds 1000000 links to projection network; while a single author paper only a loop. For this reason, the number $co_A[a,b]$ is not the best measure for measuring the collaboration intensity.
\end{itemize}

The case of collaboration between countries is slightly different because the two-mode network $WC$ is weighted. Actually, we could get it as $\mathbf{WC} = \mathbf{WA} \cdot \mathbf{AC}$ where $AC$ is the author-to-country affiliation network. This view opens a possibility to deal with authors affiliated to different countries provided that $\sum_c ac[a,c] = 1$. If the affiliations are changing through time the temporal quantities can be used \cite{tempbib}.

To obtain a collaboration network between a set of countries $C$ based on a set of works $W$, we start with a two-mode network $WC$ described by a matrix $\mathbf{WC} = [wc[w,c]]$ where
\[ wc[w,c] = \mbox{number of authors of the work $w$ from the country $c$} \]
In the network $WC$ we can consider all authors of selected works $W$ by adding to the set of countries $C$ also the "country" Others. Instead of countries other partitions of the set of authors can be used, for example institutions.

We will use $T(N) = \sum_{e \in L} w(e)$ to denote the total sum of weights of all links of the network $N=(V,L,w)$.

\section{Collaboration counting network}

The \keyw{authors counting collaboration} network $Co_C$ described by the matrix $\mathbf{Co_C}$ is obtained by projection
\[ \mathbf{Co_C} = \mathbf{WC}^T \cdot \mbox{bin}(\mathbf{WC}) \] 
where $\mbox{bin}(\mathbf{WC}) = [\widehat{wc}[w,c]] $, and $\widehat{wc}[w,c] = 1$ iff $wc[w,c] \ne 0$ and $0$ otherwise. 

What are the meaning of the entry $co_C[a,b]$ and their properties?

\renewcommand{\theenumi}{\textbf{\alph{enumi}}}
\begin{enumerate}
\item For $a\ne b$, $co_C[a,b] = \sum_w wc[w,a] \cdot \widehat{wc}[w,b]$ -- number of appearances of \emph{\textbf{authors}} from the country $a$ in works co-authored also by some author from the country $b$. We will denote this number $\mbox{wdeg}_{WC}(a/b)$.
\item $co_C[a,a] =  \sum_w wc[w,a] \cdot \widehat{wc}[w,a] = \mbox{wdeg}_{WC}(a)$ --  number of appearances of authors from the country $a$ in works from $W$; a column sum for country $a$ in the matrix $\mathbf{WC}$.
\item  From a simple example 
\[
  \mathbf{WC} = \kbordermatrix{
    & c_1 & c_2 & c_3 \\
    w_1 & 0 & 2 & 1 \\
    w_2 & 2 & 1 & 0 \\
    w_3 & 1 & 3 & 1 \\
    w_4 & 3 & 0 & 2 \\
    w_5 & 2 & 3 & 1 \\
    w_6 & 1 & 0 & 3
  }
\qquad
  \mathbf{Co_C} = \kbordermatrix{
    & c_1 & c_2 & c_3 \\
    c_1 & 9 & 5 & 7 \\
    c_2 & 7 & 9 & 8 \\
    c_3 & 7 & 3 & 8 
  }
\]
we see that the matrix $\mathbf{Co_C}$ is in general not symmetric -- there can exist pairs $a, b$ such that $co_C[a,b] \ne co_C[b,a]$.
\item Consider a row sum $R(a)$ for the country $a$ in the matrix  $\mathbf{Co_C}$. We get
\[ R(a) = \sum_b co_C[a,b] = \sum_w wc[w,a] \cdot \sum_b \widehat{wc}[w,b] = \sum_w wc[w,a] \cdot \mbox{deg}_{WC}(w) \]
Since in the network $WC$ only works $W$ with co-authors from at least 2 countries are considered, we have $\mbox{deg}_{WC}(w) \geq 2$ and we can continue
\[ R(a) \geq 2 \sum_w wc[w,a]  = 2\ \mbox{wdeg}_{WC}(a) \]
Now, combined with \textbf{b}, we finally get
\[  \sum_{b: b \ne a} co_C[a,b] \geq  \mbox{wdeg}_{WC}(a)  = co_C[a,a] \]
The sum of the out-diagonal entries in the $a$ row of the matrix $\mathbf{Co_C}$ is larger or equal to its diagonal entry.

From the example in \textbf{c} we see that this property does not hold for columns -- see the column $c_2$. 
\item For the diagonal values of the network $Co_C$ it holds $co_C[c,c] = \mbox{wdeg}_{WC}(c)$
\[ co_C[c,c] = \sum_w wc[w,c] \cdot \widehat{wc}[w,c] = \sum_w wc[w,c] = \mbox{wdeg}_{WC}(c) \]
Therefore $\sum_c  co_C[c,c] = T(WC)$.
\item In the case when also the matrix $\mathbf{WC}$ is binary, $ \mbox{bin}(\mathbf{WC}) = \mathbf{WC}$, we deal with the standard projection mentioned in the introduction $ \mathbf{Co_C} = \mathbf{WC}^T \cdot \mathbf{WC}$. In the \keyw{works counting collaboration} network $ \mathbf{Co_b} = \mbox{bin}(\mathbf{WC})^T \cdot \mbox{bin}(\mathbf{WC})$ its weight $co_b[a,b]$ counts \emph{\textbf{works}}:
$ co_b[a,b] = $ number of works from $W$ co-authored by authors from countries $a$ and $b$, and $ co_b[a,a] = $ number of works from $W$ co-authored by authors from the country $a$. 
Note that the inequality from \textbf{d} still holds (and also for columns).

\end{enumerate}

\section{Fractional approach}

For binary networks, we define their normalized versions: \keyw{standard} $n(\mathbf{WA}) = [ wan[w,a]]$
\[ wan[w,a] = \frac{wa[w,a]}{\max(1,\deg_{W\!\!A}(w))} \]
and \keyw{strict} (or Newman's) $N(\mathbf{WA}) = [ waN[w,a]]$
\[ waN[w,a] = \frac{wa[w,a]}{\max(1,\deg_{W\!\!A}(w)-1)} \]
Using the normalized networks we define the \keyw{standard fractional projection}
\[ \mathbf{Co_n} = n(\mathbf{WA})^T \cdot n(\mathbf{WA}) \]
and the \keyw{strict fractional projection}
\[ \mathbf{Co_N} = D_0(n(\mathbf{WA})^T \cdot N(\mathbf{WA})) \]
where the function $D_0(\mathbf{M})$ sets the diagonal of a square matrix  $\mathbf{M}$ to $0$.

We know \cite{fraca} that if $\deg_{W\!\!A}(w) > 0$, each work $w \in W$ contributes equally, a unit 1, to the total weight of links in $ \mathbf{Co_n}$. The same holds for  $\mathbf{Co_N}$ if $\deg_{W\!\!A}(w) > 1$.

To extend the fractional projections to weighted two-mode networks we define for the \keyw{standard} case $n(\mathbf{WC}) = [ wcn[w,c]]$
\[ wcn[w,c] = \frac{wc[w,c]}{\max(1,\mbox{wdeg}_{WC}(w))} \]
Again we have $T(Co_n) = |W|$ for $ \mathbf{Co_n} = n(\mathbf{WC})^T \cdot n(\mathbf{WC}) $.
\[
  \mathbf{Co_b} = \kbordermatrix{
    & c_1 & c_2 & c_3 \\
    c_1 & 5 & 3 & 4 \\
    c_2 & 3 & 4 & 3 \\
    c_3 & 4 & 3 & 5 
  }
\qquad
  \mathbf{Co_n} = \kbordermatrix{
    & c_1 & c_2 & c_3 \\
    c_1 & 1.0180556 & 0.5088889 & 0.5230556 \\
    c_2 & 0.5088889 & 1.1655556 & 0.4255556 \\
    c_3 & 0.5230556 & 0.4255556 & 0.9013889 
  }
\]

There is no obvious way how to define the strict normalization for weighted networks.

There is another possible view on fractional projections. The definition of matrix $n(\mathbf{WA})$ can be written as $n(\mathbf{WA}) = \mathbf{d_n} \cdot \mathbf{WA}$ and similarly $N(\mathbf{WA}) = \mathbf{d_N} \cdot \mathbf{WA}$ where $\mathbf{d_n}$ is a diagonal $W \times W$ matrix with $d_n[w,w] = 1/\max(1,\deg_{W\!\!A}(w))$ and $\mathbf{d_N}$ with $d_N[w,w] = 1/\max(1,\deg_{W\!\!A}(w)-1)$.

In both cases we get ($\mathbf{d}^T = \mathbf{d}$)
\[  \mathbf{Co_n} = n(\mathbf{WA})^T \cdot n(\mathbf{WA})  = \mathbf{WA}^T \cdot \mathbf{d_n}\cdot \mathbf{d_n} \cdot \mathbf{WA} \]
\[  \mathbf{Co_N} = n(\mathbf{WA})^T \cdot N(\mathbf{WA}) = \mathbf{WA}^T \cdot \mathbf{d_n}\cdot \mathbf{d_N} \cdot \mathbf{WA} \]
Because a product of diagonal matrices is a diagonal matrix, $\mbox{diag}(a_w) \cdot \mbox{diag}(b_w) = \mbox{diag}(a_w \cdot b_w)$, both cases have a common form $\mathbf{WA}^T \cdot \mathbf{d} \cdot \mathbf{WA}$. It can be related to the weighted scalar product. Maybe this form can lead also to the extension of strict projection for weighted two-mode networks.

\section{Strict fractional collaboration}

Let us look at a simple example. Assume, that a work $w$ has authors from three countries $a$, $b$, and $c$. Then, since the co-authors inside the same country do not count, its contribution $ T(w)$ to the total weight, see the contribution matrix 
\[   \mathbf{Co_C}(w) = \kbordermatrix{
    & a & b & c \\
    a & \mathbf{0} & wc[w,a] \cdot wc[w,b] & wc[w,a] \cdot wc[w,c] \\
    b & wc[w,b] \cdot wc[w,a] & \mathbf{0} & wc[w,b] \cdot wc[w,c] \\
    c & wc[w,c] \cdot wc[w,a] & wc[w,c] \cdot wc[w,b] & \mathbf{0}
  },
\]
is $T(w) = \sum_{e,f \in \{a,b.c\} \land e \ne f} wc[w,e] \cdot wc[w,f] $. By the rule of product and the rule of sum from basic combinatorics \cite{rules}, $T(W)$ is equal to twice the number of all co-authorships of authors from different countries -- pairs $(a,b)$ and $(b,a)$ are representing co-authorship of authors $a$ and $b$.

To make $T_N(w) = 1$ we must set the entry $d_N[w,w]$ of
the diagonal matrix $\mathbf{d_N}$ for the weighted network $\mathbf{WC}$ to $d_N[w,w] = 1/T(w) = 1/(\mbox{wdeg}_{WC}(w)^2 - \sum_c wc[w,c]^2)$. Note that $\sum_c wc[w,c] = \mbox{wdeg}_{WC}(w)$ and
\[ \mbox{wdeg}_{WC}(w)^2 - \sum_c wc[w,c]^2 = \sum_{e,f : e \ne f} wc[w,e] \cdot wc[w,f] \] 
The left side of this equality is computationally more convenient.

{

\begin{algorithm}
\caption{Computing projection matrices $\mathbf{Co_b}$, $\mathbf{Co_n}$, $\mathbf{Co_C}$, and $\mathbf{Co_N}$.\label{proj}}
\begin{algorithmic}[1]
\Function{Projections}{$\mathcal{W}$,$C$}
\State $Cob \gets Con \gets CoC \gets CoN \gets matrix(0, nrow=|C|,ncol=|C|)$
\For{$w \in \mathcal{W}$}
   \State determine $wc[w,c]$ for $c \in C$
   \State $Cw \gets \{ c \in C: wc[w,c] > 0 \}$
   \State $wdegw \gets \sum_{c \in Cw} wc[w,c]$
   \State $sqw \gets \sum_{c \in Cw} wc[w,c]^2$
   \State $dnw \gets 1/wdegw^2$
   \State $dNw \gets 1/(wdegw^2 - sqw)$
   \For{$e \in Cw$}
      \For{$f \in Cw$}
         \State $Cob[e,f] \gets Cob[e,f] + 1$
         \State $CoC[e,f] \gets CoC[e,f] + wc[w,e]$
         \State $Con[e,f] \gets Con[e,f] + wc[w,e] \cdot wc[w,f] \cdot dnw$
         \If{$e \ne f$}
            \State $CoN[e,f] \gets CoN[e,f] + wc[w,e] \cdot wc[w,f] \cdot dNw$
         \EndIf
      \EndFor  
   \EndFor
\EndFor
\State \Return $Cob$, $Con$, $CoC$, $CoN$

\EndFunction

\end{algorithmic}
\end{algorithm}
}

It is easy to see that we made a good guess -- in the corresponding projection $\mathbf{Co_N} = D_0(\mathbf{WC}^T \cdot \mathbf{d_N} \cdot \mathbf{WC})$  each work contributes equally, a unit 1, to the total of link weights.
\[ T_N(w)  = d_N[w,w] \cdot \sum_{e,f : e \ne f} wc[w,e] \cdot wc[w,f] = 1 \]
Therefore
\[ T(Co_N) = \sum_w T_N(w) = |W| \]
For our example from Section 2 \textbf{c} we get
\[
  \mathbf{Co_N} = \kbordermatrix{
    & c_1 & c_2 & c_3 \\
    c_1 & 0.000000 & 0.9870130 & 1.1623377 \\
    c_2 & 0.987013 & 0.0000000 & 0.8506494 \\
    c_3 & 1.162338 & 0.8506494 & 0.0000000 
  }
\]
with $T(Co_N) = 6$.

\section{Computing}
$C$ is a set of countries of our interest.
$\mathcal{W}$ is a list of metadata about the works from the selected bibliographic data source, co-authored by authors from at least two different countries from $C$. All four projection matrices $\mathbf{Co_b}$, $\mathbf{Co_n}$, $\mathbf{Co_C}$, and $\mathbf{Co_N}$ can be constructed in a single run through the list using the Algorithm \ref{proj}.

Notes on the implementation of the algorithm:
\begin{itemize}
\item If we do not need the network $WC$ we essentially need in line 4 the current list of pairs $(c,wc(c))$ for $wc(c) > 0$.
\item Networks $Co_b$, $Co_n$, and $Co_N$ are symmetric. They can be represented by an undirected network with the weight of an edge $(e:f)$ equal to twice the computed value, except for loops. The computation can be restricted to pairs $(e,f)$ for which $e \leq f$.
\end{itemize}

\section{Conclusions}

In the paper, we derived the results in terms of the binary authorship network $W\!\!A$ and the weighted network $W\!C$. The results hold in general for similar weighted two-mode networks such as (journals, universities, number of published articles of authors from the university $u$ in the journal $j$ in the selected time interval), (territorial units, universities, number of students from the territorial unit $t$ studying this year at the university $u$), (web resources (movies or music tracks), types of resource, number of times the resource $r$ of type $t$ was downloaded in the selected time interval), (retail chain customers (chain card owners), (types of) products, the value of the product $p$ bought by the customer $c$ in the selected time interval),  etc.

An application of the proposed projections in an analysis of a large real-life data set will be published in a separate paper(s).

\section*{Acknowledgments}

\ifblind \else
This work is supported in part by the Slovenian Research Agency
 (research program P1-0294 and research projects J5-2557, J1-2481 and J5-4596),
 and prepared within the framework of the COST action CA21163 (HiTEc).
\fi




\begin{thebibliography}{99}

\bibitem{fraca}
Batagelj, V: On fractional approach to analysis of linked networks. Scientometrics 123 (2020) 2: 621-633 
\bibitem{onbib}
Batagelj, V, Cerinšek, M: On bibliographic networks. Scientometrics 96 (2013) 3, 845-864.
\bibitem{tempbib}
Batagelj, V, Maltseva, D: Temporal bibliographic networks. Journal of Informetrics, Volume 14, Issue 1, February 2020, 101006.
\bibitem{soviet}
Matveeva, N., Batagelj, V., Ferligoj, A.: Scientific collaboration of post-Soviet countries: the effects of different network normalizations. Scientometrics 128 (2023), 4219–4242.
\bibitem{rules} Wikipedia: Combinatorial principles.\\
\href{https://en.wikipedia.org/wiki/Combinatorial_principles}{https://en.wikipedia.org/wiki/Combinatorial\_principles}
\end{thebibliography}
\end{document}